\documentstyle[preprint,aps]{revtex}

\def\a{\alpha}
\def\b{\beta}
\def\c{\chi}
\def\d{\delta} 			\def\D{\Delta}
\def\e{\epsilon}

\def\g{\gamma} 			\def\G{\Gamma}

\def\l{\lambda} 		\def\L{\Lambda}
\def\m{\mu}
\def\n{\nu}
 			
\def\p{\pi}

\def\fr{\frac}
\def\ba{\begin{array}}
\def\ea{\end{array}}
\def\bz{\begin{equation}}
\def\ez{\end{equation}}
\def\by{\begin{eqnarray}}
\def\ey{\end{eqnarray}}

\def\nn{\nonumber}

\newtoks\slashfraction
	\slashfraction={.13}
\def\slash#1{\setbox0\hbox{$\, #1$}
	\setbox0\hbox to \the\slashfraction\wd0{\hss \box0}/\box0}

\begin{document}
\tightenlines
\draft

\preprint{\vbox{\hbox{FERMILAB-Pub-94/120-T}
               \hbox{LA-UR-94-1637}
                \hbox{ZU-TH-10/94}
                \hbox{May 1994}
                \hbox{T/E} }}
\vskip 0.5truecm

\title{Radiative Leptonic Decays of Heavy Mesons}
\author{Gustavo Burdman}
\address{Fermi National Accelerator Laboratory, Batavia, Illinois  
60510}

\author{T.\ Goldman}
\address{Los Alamos National Laboratory, Los Alamos, New Mexico  
87545}

\author{Daniel Wyler}
\address{Institut f\"ur Theoretische Physik, Universit\"at Z\"urich, Z\"urich, 
Switzerland}

\maketitle

\begin{abstract}
We compute the photon spectrum and the rate for the decays $B(D)\to
l\nu_l\gamma$ These photonic modes constitute a potentially large
background for the purely leptonic decays  which 
are used to extract the heavy
meson decay constants. While the rate for
$D\to l\n\g$ is small, the
radiative decay in the $B$ meson case could be of comparable
magnitude or even larger than $B\to\m\n$. This would affect the determination
of $f_B$  if the $\tau$ channel cannot be identified. 
We obtain theoretical
estimates for the photonic rates and disscuss their possible
experimental implications.
\end{abstract}

%\pacs{12.39.Hg, 12.39.Jh, 13.20.He, 13.20.Fc}

\vskip 0.5truecm
%\centerline{(Submitted to Physical Review Letters)}
\newpage

\narrowtext
\section{Introduction}
\label{sec:1}
The leptonic decays of heavy mesons are of great interest both
theoretically and experimentally. The purely leptonic decay $B^-\to
l^-\nu_{l}$ can be used to extract $|V_{ub}|$ by using predicted values
for the $B$ meson decay constant $f_B$ from lattice calculations. Or
conversely, if $|V_{ub}|$ is measured in semileptonic decays, the
leptonic decay is in principle the only mode to access $f_B$. On the
other hand, the situation in the charmed mesons is clearer given our
knowledge of the CKM mixing angles involved ($|V_{cs}|$ and
$|V_{cd}|$). Recently the observation  of the decay $D_s\to\m\n_{\m}$
has been reported \cite{fds}.  Although the results are still
preliminary it is a first step towards understanding the behavior of
heavy meson decay constants. The experimental difficulty in the
measurement of purely leptonic decays of heavy pseudoscalars is due
mostly to the well known effect of helicity suppression: back-to-back
leptons must make a spin $0$ final state, but the anti-neutrino is
right-handed and forces the charged lepton to this helicity which
introduces a factor of the lepton mass in the amplitude. In the end the
decay rate is suppressed by the factor $(m_l/m_H)^2$, where $m_H$ is
the heavy meson mass. For instance for the $B$ meson
\bz
\G(B\to l\bar{\n}_l)
=\fr{G_{F}^{2}}{8\p}|V_{ub}|^2f_{B}^{2}\left(\fr{m_l}{m_B}\right
)^2m_{B}^{3}\left(1-\fr{m_{l}^{2}}{m_{B}^{2}}\right) \label{leprate}
\ez
where the pseudoscalar meson decay constant is defined by
\bz
\langle 0|\bar{q}\g_{\m}(1-\g_5)b|B(P)\rangle =if_BP_\m \label{psdc}
\ez
and analogous expressions can be written for $D$ mesons.  Thus, when
the charged lepton is an electron the purely leptonic decay is
practically inaccessible. At the other extreme, when the charged
lepton is a tau there is no suppression. However the
observation of this decay is experimentally difficult (only one
visible particle in the final state). Muons seem more promising to allow the
observation of the decay constants of $D$ and $B$ mesons. However in
$B\to\mu\nu$ the suppression factor is still $4\times 10^{-4}$ giving a
branching fraction of $10^{-7}-10^{-6}$.

There are other decays that indirectly involve the heavy meson decay
constants. For instance, the decay $B\to\p l\nu_l$ is expected to be
largely dominated by a $B^*$ pole diagram at very low recoiling pion
energies \cite{iw1,wise,bd}. This implies the presence of the
vector-meson decay constant $f_{B^*}$ which can be related to $f_B$ by
Heavy Quark Spin Symmetry \cite{hqs}. However this region of phase
space is difficult to access due to kinematic suppression.

In this letter we investigate the decay modes $B^-(D^-)\to
l^-\nu_l\gamma$.  There are two types of contributions: Internal
Bremsstrahlung (IB) and Structure Dependent (SD) photon emission 
\cite{goldman}.
 As is known the IB  contributions are still helicity suppressed. On
the other hand, the SD contributions are reduced by  the
electromagnetic coupling constant $\alpha$ but they are  {\it not} 
suppressed
by the charged lepton mass. Therefore, what in principle could be
regarded as a mere radiative correction to the purely leptonic decays
has the potential to be of comparable magnitude and in some cases even
much larger. In what follows we analyze the situation in charmed
and beauty mesons. In Sec.~2 we establish the phenomenological
relevance of these decays and in Sec.~3 we discuss theoretical
estimates of the unknown constants involved. Conclusions and comments
are presented in Sec.~4.

\section{Phenomenology of $H\to\mu\nu\g$}
\label{sec:2}
As mentioned in the previous section, the $\mu$ modes are the most
interesting from the point of view of the extraction of heavy meson
decay constants. We will concentrate on the case $l=\mu$ but the
treatment for $l=e$ is analogous and the numerical differences between
these two cases will be stressed when relevant.

The emission of a real photon in leptonic decays of heavy mesons can
proceed via the two mechanisms mentioned in Sec.~\ref{sec:1}. The
possible IB diagrams are shown in Fig.~1. The  corresponding amplitude
is given by
\by
{\cal M}_{PB}&=&m_\mu f_B\times\left\{\left(\fr{\e\cdot p_l}{p_l\cdot
k}-F\fr{\e\cdot P}{P\cdot k}\right)\bar{\mu}(1-\g_5)\nu\right. \nn \\
& &\left. +\fr{1}{2p_l\cdot k}\bar{\mu}\slash\e\slash
k(1-\g_5)\nu\right\}\label{mpb}
\ey
where $P$, $p_l$ and $k$ are the four-momenta of the decaying meson,
the charged lepton and the emitted photon, respectively, $\e$ is the
polarization of the photon and $F$ is the electromagnetic form-factor
of the $B$ meson. (This is not simply related to the Isgur-Wise
function \cite{iwfunc}
given that the matrix element of the electromagnetic current
between two $B$ mesons receives contributions from both the heavy and
light quarks). The important feature of (\ref{mpb}) is that it is
suppressed by a factor of the lepton mass, as one would expect of
bremsstrahlung photons.

On the other hand, the SD diagrams in Fig.~2 involve the
contributions from heavy intermediate states coupling to the initial
heavy pseudoscalar and the photon. In Fig.~2 we show the contributions
from vector and axial-vector mesons. The helicity suppression is
avoided because  the meson directly coupling to the lepton
pair has spin one.  These types of diagrams were previously considered
in the context of light pseudoscalar decays, in particular in $\p\to
l\nu\g$ and $K\to l\nu\g$ \cite{goldman}. It was also shown there that this
phenomenological picture in fact correctly accounts for the SD contributions.

The general form of the
amplitude corresponding to Fig.~2(a) requires the knowledge of  the
$B^*B\g$ coupling. This is defined in the process $B^*\to B\g$ with the
amplitude given by
\bz
{\cal A}(B^*\to B\g)=e\m_V\e^{\a\b\g\d}\e^{*}_{\a}v_\b k_\g\l_\d
\label{bbg}
\ez
where $v$ is the velocity four-vector of the decaying particle, $k$ is
the photon four-momentum, $\l$ is the polarization four-vector of the
$B^*$ and $\m_V$ is a constant characterizing the strength of the M1
transition.  Eqn.~(\ref{bbg}) can be used to write the amplitude of
Fig.~2(a) by having first the $B$ meson decaying into a real photon and an
off-shell vector meson, which after propagating decays weakly through
\bz
\langle 0|\bar{q}\g_\m(1-\g_5)b|B^*(v,\l)\rangle =if_{B^*}\e_\m
\label{vmdc}
\ez
Eqn.~(\ref{vmdc}) defines the vector-meson decay constant. Inserting the heavy
vector-meson propagator $i(-g_{\m\n}+v_\m v_\n)/2(v.k+\D)$, where
$\D=m_{B^*}-m_B$, we obtain its contribution to the SD diagram
\bz
{\cal
A}_V(B\to\m\n\g)=e\m_{V}\fr{1}{2(v.k+\D)}f_{B^*}\e^{\a\b\g\d}\e^
*_{\a}v_\b k_\g {\cal L}_\d \label{vecamp}
\ez
with ${\cal L}_\d=\bar{u}_l\g_\d (1-\g_5)v_\n$ the lepton current.

There will also be contributions  from heavy  axial-vector meson 
states ($J^P=1^+$). The Heavy Quark
Symmetries \cite{hqs} of the strong interactions and their consequences
in the hadron spectrum will help us to identify the relevant
contributions.  In the Heavy Quark Limit the spin of the heavy quark
decouples from the light degrees of freedom. Thus their total angular
momentum $j_l$ can be used as a good label. The meson angular momentum
is $J=1/2\pm j_l$. For the ground state $j_l=s_l$, the spin carried by
the light degrees of freedom and therefore a $(0^-,1^-)$ spin doublet
is predicted. This can be identified with the $(B,B^*)$. If we allow
for orbital angular momentum $L=1$, then there will be two additional
spin doublets as excited states. They correspond to $j_l=1/2$ and
$j_l=3/2$ and their parity is even: $(0^+,1^+)$ and $(1^+,2^+)$. Thus
in principle there could be two different axial-vector mesons
contributing to $B\to\m\n\g$.  We write down the amplitude for the
process in Fig.~2(b), which is the generic  contribution of an
axial-vector meson intermediate state.   The coupling is of the form
\bz
{\cal A}_{A}^{(i)}(B_1\to B\g)=e\m_{A}^{i}(v\cdot  
k\l\cdot\e-v\cdot\e\l\cdot
k) \label{axial}
\ez
where the superscript $i=1/2,3/2$ identifies the spin-parity doublet to
which the axial-vector meson belongs.  The axial-vector decay constants
$f_{A}^{1/2}$ and $f_{A}^{3/2}$ are defined analogously to $f_B$ in
(\ref{vmdc}). Putting these together the contributions of Fig.~2(b)
take the form
\bz
{\cal A}_{A}^{(i)}(B\to\m\n\g)=\fr{e\m_{A}^{(i)}f_{A}^{i}}{2(v\cdot
k+\D_i)}\left
[v\cdot\e k_\m-v\cdot k\e_\m\right ]{\cal L^{\mu}} \label{axiamp}
\ez
where $\D_{1/2}=m_{B_1'}-m_B$ and  $\D_{3/2}=m_{B_1}-m_B$ are the
excitation energies of the $j_l=1/2$ and $j_l=3/2$ even parity doublets
and we again neglected pieces proportional to the charged lepton mass.
It is worth noticing that the axial-vector pseudoscalar mass
differences $\D_{1/2}$ and $\D_{3/2}$ are quantities that remain
constant in the limit $m_Q\to\infty$ whereas $\D$ goes to zero in the
same limit. In the Heavy Quark Effective Theory (HQET) \cite{hqet} a
term that goes as $\D^2$ can be neglected in the $B^*$ propagator on
the basis of the heavy mass suppression. In the case of the $B_1$ or
$B_1'$ propagators in Fig.~2(b) the quadratic terms in $\D_{1/2}$ and
$\D_{3/2}$ can still be considered numerically small even when they
are  quantities of order $1$ in the heavy masses, when compared with
$2m_B$. 

Most importantly,  Heavy Quark Spin Symmetry can be used to obtain
 relations
between the decay constants of particles in the same parity multiplet.
For instance
\bz
f_{B^*}=m_Bf_B \label{bbstar}
\ez
and analogous relations can be written between the decay constants of
the members of the $j_{l}^{P}=1/2^+$  and  the $j_{l}^{P}=3/2^+$
doublets. However HQS does not predict relations among decay constants
of members of different spin-parity doublets. Then the contributions
from excited states will imply the presence of new unknowns other than
the pseudoscalar decay constant. Defining $x=2E_\g/m_B$ and
$y=2E_l/m_B$ as the rescaled photon and charged lepton energies in the
$B$ rest frame  the double differential decay rate is given by
\by
\fr{d^2\G}{dxdy}&=&\fr{G_{F}^{2}|V_{ub}|^2}{32\p^2}\a
m_{B}^{3}f_{B^*}^{2}\m_{V}^2
\times
\left\{\fr{1}{(x+\fr{2\D}{m_B})^2}+\left(\sum_{i}\fr{\g_{i}
}{x+\fr{2\D_i}{m_B}}\right)^2\right\} \nn \\
& &\times\left[y^2(1-x)+y(3x-x^2-2)+\fr{1}{2}(3x^2-4x-x^3+2)\right]
 \label{phospec}
\ey
where we defined
\bz
\g_i=\fr{\m_{A}^{(i)}f_{A}^{i}}{\m_{V}f_{B^*}} \label{gamma}
\ez
as the relative axial-vector to vector meson coulping strength. 
 The fact that there are no relations among  decay constants of states
of different spin-parity doublets will be reflected in the persistence
of the unknown $\g_i$'s when we normalize to $\G(B\to\m\n)$. We will
analyze the $\g_i$'s and $\m_{V}$ from the theoretical point of view in
the next section. For now let us assume that  $\m_V$ and
$\m_{A}^{(i)}$  do not depend on the photon energy so we can integrate
and compare with the purely leptonic decay. The result is
\by
R_{B}^{\m}&=&\fr{\G(B\to\m\n\g)}{\G(B\to\m\n)}\nn \\
&=&\fr{1}{6\p}\a \m_{V}^2 m_{B}^{2}\left(\fr{m_B}{m_\m}\right)^2
\times\int_{0}^{1}dx
x^3(1-x)\left\{\fr{1}{(x+\fr{2\D}{m_B})^2}+\left(\sum_{i}\fr{\g_{i}
}{x+\fr{2\D_i}{m_B}}\right)^2\right\} \label{ratiomu}
\ey
To have an idea of the potential importance of the photonic mode we
take a definite value for the mass differences $\D_i=600$~MeV as
suggested by the charmed meson system \cite{pdg}. This gives  
approximately
\bz
R_{B}^{\m}\approx  
2\m_{V}^2\left(1+\fr{(\g_{1/2}+\g_{3/2})^2}{2}\right) \text{GeV}^2
\label{estimu}
\ez
which shows that unless there are unnaturally small $B$ photon
couplings (see next section) the photonic decay, which in principle
could have been considered a small radiative correction, will dominate
the leptonic decay or will be at least of comparable magnitude.
Therefore it is important to have a good theoretical understanding of
the couplings   of all the relevant intermediate states in order to
substract these events as a background for $B\to\m\n$.  The fact that,
to this order in $(m_\m /m_B)^2$, the result of Eqn.~(\ref{phospec}) is
independent of the lepton mass implies that
\bz
\G(B\to e\n\g)=\G(B\to\m\n\g) \label{mueqel}
\ez
and that
\bz
\G(B\to e\n\g)\gg\G(B\to e\n)
\ez
allows for the separation of both effects by using the electronic modes
as well. Thus integrating (\ref{phospec}) over the photon energy and
integrating the resulting $\m$ spectrum around the end point over a
region the size of the experimental $\m$ energy resolution in
$B\to\m\n$ will eliminate the background. Perhaps even more
interesting, given the richness of the physics that enters in them, is
the possibility of observing the photonic modes at branching ratios that
will soon be accesible ($\approx 10^{-5}-10^{-6}$).

The treatement of  charmed mesons is entirely analogous to $B$ mesons.
With the obvious replacements in (\ref{ratiomu}) we obtain
\bz
R_{D}^{\m}\approx  4\times
10^{-2}\m_{V}^{2}\left(1+\fr{(\g_{1/2}+\g_{3/2})^2}{2}\right) \text{GeV}^2 
\label{estimu2}
\ez
which is suppressed by the factor $(m_D/m_B)^4$ in rescaling
(\ref{ratiomu}).  Therefore the effect is, as expected, less spectacular
in the $D$ mesons although of a branching fraction comparable to that
of the effect in the $B$ system.

In the following section we discuss various theoretical estimates of
$\m_{V}$ and the $\g_i$'s.

\section{Theoretical Estimates}
\label{sec:3}
The radiative leptonic decays depend crucially on
the vector and axial-vector couplings to the heavy pseudoscalar, $\m_V$
and $\m_{A}^{(i)}$ as well as the ratio of the axial-vector to vector
meson decay constants.   The photon couples to both the heavy and light
quark pieces of the electromagnetic current.  Heavy Quark Symmetry
(HQS) fixes the heavy quark contribution. In the Heavy Quark Effective
Theory \cite{hqet} the coupling of the heavy quark $Q$ to the
electromagnetic field responsible for the $B^*$ to $B$ transition is
given by the operator
\bz
\fr{Q_Q}{2m_Q}\bar{h}_v\sigma_{\m\n}h_vF^{\m\n}
\ez
where $h_v$ is the heavy quark field charaterized by its four-velocity
$v$ and $Q_Q$ is its electromagnetic charge. This leads to
\bz
\m^{(h)}=\fr{Q_Q}{m_Q}
\ez
where the superscript $h$ indicates a contribution from the heavy
quark.

The piece of the $B^*B\g$ coming from the coupling to the light degrees
of freedom is not given by HQS. In the $SU(3)$ limit can be written as
\cite{amud}
\bz
\m^{(l)}=Q_q\b \label{mulite1}
\ez
where $Q_q$ is the charge of the light degrees of freedom and $\b$ is
an unknown quantity. In the non-relativistic quark model this is
predicted to be
\bz
\b_{NRQM}=1/m_q
\ez
with $m_q$ the constituent light quark mass ($m_u\approx m_d\approx
330$~MeV and $m_s\approx 450$~MeV). On more general grounds $\b$ is
expected to be of the order $1/\L_{QCD}$. If we take the HQS  and the
NRQM predictions for $\m^{(h)}$ and $\m^{(l)}$ respectively in order
to have an estimate of the magnitude of the coupling we have the
simple expression
\bz
\m_V=\fr{Q_Q}{m_Q}+\fr{Q_q}{m_q} \label{mutot1}
\ez
which is the naive quark model result but now supplemented by HQS.  In
the charmed mesons, it is sufficient to explain the ratio $\G(D^{*0}\to
D^0\g)/\G(D^{*+}\to D^+\g)$ \cite{cleo1}.  Therefore we will rely on
(\ref{mutot1}) for estimates in the $B$ meson decays. For $m_b=5$~GeV
and taking $\b=3 \text{GeV}^{-1}$, we see that the value for the coupling 
\bz
\m_V\approx -2GeV^{-1} \label{qmres}
\ez
is such that the ratio (\ref{ratiomu}) could be sizeable.

The axial-vector meson couplings are expected to be of the same order
as $\m_V$. In fact the NRQM gives $\m_{A}^{(1/2)}=\m_V/\sqrt{3}$ and
$\m_{A}^{(3/2)}=\sqrt{2/3}\m_V$, where the superscripts indicate an
axial-vector meson belonging to the $(0^+,1^+)$ and $(1^+,2^+)$
doublets repectively.  The contributions of Fig.~2(b) are also
proportional to $f_{A}^{i}$, the axial-vector meson decay constants.
But in the NRQM these are zero given that the wave function at the
origin for an orbitally excited state vanishes due to the presence of a
centrifugal barrier. Although this seems to imply that  $\g_i\ll 1$,
relativistic effects could be important and modify this prediction
drastically.  Unfortunately there is no conclusive experimental
indication from other hadronic systems about the size of the
axial-vector meson contributions. In experiments involving
$\p\to\m\n\g$ and $K\to\m\n\g$ \cite{pdg} the ratio of the axial-vector
to vector contributions is consistent with zero and also with being
large.

In order to address corrections to this simple picture it is useful to
consider an effective theory coupling heavy hadrons (in this case
mesons) to goldstone bosons and also low energy photons. This theory
incorporates both Heavy Quark Symmetry and Chiral Symmetry by
introducing the spin-parity doublet in the form of $4\times 4$ matrices
as follows \cite{wise,bd,yan,falk}
\begin{mathletters}
\label{states}
\bz
H_a=\fr{(1+\slash v)}{2}\left\{\slash B^*-B\g_5\right\} \label{grst}
\ez
\bz
S_a=\fr{(1+\slash v)}{2}\left\{\slash B_{1}'\g_5-B_{0}^{*}\right\}  
\label{exc1}
\ez
\bz
T_{a}^{\m}=\fr{(1+\slash 
v)}{2}\left\{B^{*\m\n}_{2}\g_\n-\sqrt{\fr{2}{3}}B^{\n}_{1}\g_5(g^{\m}_ 
{\n}-\fr{1}{3}\g_\n (g^\m-v^\m))\right\} \label{exc2}
\ez
\end{mathletters}
\noindent where $a=1,2,3$ is the $SU(3)$ index of the light degrees of
freedom in the hadron. $H_a$ corresponds to the $(0^-,1^-)$ ground
state doublet; $S_a$ is the excited state doublet $(0^+,1^+)$
corresponding to the spin of the light degrees of freedom being
$j_l=1/2$ and $T_{a}^{\m}$ is the excited state doublet $(1^+,2^+)$
corresponding to $j_l=3/2$.

An important set of corrections are those arising from the loop
diagrams of Fig.~3, where the photon couples to a goldstone boson in
the loop.  This type of correction therefore is not suppressed by the
heavy mass. They were first considered in Ref.~\cite{amud} for the
$D^*D\g$ coupling in the context of Heavy Hadron Chiral Perturbation
Theory (HHChPTh) and introduce a non-analytic dependence on the quark
masses, $m_{q}^{1/2}$  as well as on the mass differences $\D$ and
$\D_i$ that break $SU(3)$ in (\ref{mulite1}).  To calculate these
diagrams we need the coupling of two heavy hadrons to one goldstone
boson. These are given by \cite{falk}
\by
{\cal L}_{1\p}&=&gTr\left[\bar{H}_aH_b\slash A_{ba}\g_5\right]+
g'Tr\left[\bar{S}_aS_b\slash A_{ba}\g_5\right]+
g'' Tr\left[\bar{T}_{a}^{\m}T_{\m b}\slash A_{ba}\g_5\right] \nn \\
& &+f'Tr\left[\bar{S}_aT^{\m}_{b} A_{\m ba}\g_5\right]
+f'' Tr\left[\bar{H}_aS_b\slash A_{ba}\g_5\right] \label{onegb}
\ey
with the traces taken over Dirac indices and where the first three
terms correspond to the axial couplings between members of the same
spin-parity doublet and the last two terms give the transitions between
doublets. As it was pointed out in \cite{falk} the axial couplings
between the $(1^+,2^+)$ doublet and the ground state vanish to this
order in the chiral expansion, that is
$Tr\left[\bar{H_a}T_{b}^{\m}A_{\m ba}\g_5\right]$ vanishes. This means
that the diagrams in Fig.~3 corresponding to having the
$(B_1,B_{2}^{*})$ inside the loop can be neglected. On the other hand
 we can also see from (\ref{onegb}) that to this order there will be no
$B_{1}'B\p$ coupling which prevents a contribution from the
axial-vector meson to the loop. There are four diagrams remaining which
correspond to the heavy mesons in Fig.~3(a) being (from left to right)
$(B,B^*,B^*)$; $(B,B^*,B_{1}')$; $(B,B^{*}_{0},B_1')$ and
$(B,B^{*}_{0},B_1)$. In terms of the couplings, the first one
corresponds to a correction to $\m_V$, the two
following to a correction to $\m_{A}^{(1/2)}$ coming from $B_1'\to B\g$
and the last one to a correction to $\m_{A}^{(3/2)}$ from $B_1\to
B\g$.  Finally the diagrams of Fig.~3(b)-(c) vanish in the limit
$m_l\to 0$ and will be strongly suppressed.  Therefore the only diagrams
contributing are the ones represented in Fig.~3(a), with pion and kaon
loops giving
\bz
\m_{V}=\fr{-1}{3m_b}-\fr{2}{3m_q}+\fr{g^2}{4\p^2
f_{\pi}^{2}}I(m_\p ,\D)+\fr{g^2}{4\p^2 f_{K}^{2}}I(m_K,\D)   
\label{mutot2}
\ez
where
\bz
I(m,\D)=\D\left( \ln\fr{m^2}{\m^2} +2F(m/\D)\right) \label{iloop}
\ez
with $\m$ the renormalization scale and
\bz
F(x)=\left\{\ba{c}\sqrt{x^2-1}\tanh^{-1}\sqrt{x^2-1}\qquad ;x\leq 1  
\\
-\sqrt{x^2-1}\tan^{-1}\sqrt{x^2-1} \qquad ;x\geq 1
\ea\right. \label{function}
\ez

Analogously, the corrections to $\m_{A}^{(i)}$ can be computed. With
the corrections mentioned above, the $B_{1}'B\g$ coupling is now
\bz
\m_{A}^{(1/2)}=\fr{1}{\sqrt{3}}\left(  
\fr{-1}{3m_b}-\fr{2}{3m_q}\right)+
f''(g'- g/2)\left\{\fr{1}{4\p^2 f_{\p}^{2}}I(m_\p  
,\D_{1/2})+\fr{1}{4\p^2
f_{K}^{2}}I(m_K,\D_{1/2})\right\} \label{mutotax1}
\ez
whereas the $B_1B\g$ coupling now is
\bz
\m_{A}^{(3/2)}= \fr{2}{\sqrt{3}}\left(  
\fr{-1}{3m_b}-\fr{2}{3m_q}\right)+
f'f''\left\{\fr{1}{4\p^2 f_{\p}^{2}}I(m_\p ,\D_{3/2})+\fr{1}{4\p^2
f_{K}^{2}}I(m_K,\D_{3/2})\right\} \label{mutotax2}
\ez
The quantities $g$, $g'$, $f'$ and $f''$ are expected to be of order
one on dimensional grounds. In fact the NRQM predicts $g=1$.
Unfortunately, the $D^*$ lifetime has not been measured yet, preventing
the extraction of $g$ from $D^*\to D\p$.  The experimental upper limit
\cite{guplim} is $g<0.7$. The model of Ref.~\cite{barhill} predicts
 $g=0.32$. Taking $g$  to be somewhere in between
we see that the magnetic coupling in (\ref{mutot2}) still remains a
quantity of order one or possibly larger. The same can be said about
the corrections to the contributions from axial-vector mesons, where
the corrections do not necessarily reduce the value of the couplings
given the different signs in (\ref{mutotax1}) and (\ref{mutotax2}) as
well as the potential relative phases between these strong couplings.

So far we have not made a distinction between the contributions coming
from the $B_{1}'$ axial-vector meson, corresponding to $j_l=1/2$, and
the $B_1$ corresponding to $j_l=3/2$. In general one can think that
their relative contribution is only governed by the size of
$\m_{A}^{(1/2)}$ and $\m_{A}^{(3/2)}$ and their respective decay
constants. There is already evidence from the charm meson system that
the $(0^+,1^+)$ is very broad relative to the $(1^+,2^+)$. This is
mostly due to the fact that the latter does not couple to the ground
state to leading order but only through a $D$-wave amplitude
\cite{falk}.  Therefore the narrow width approximation implicitly used
in the calculation of (\ref{phospec}) should be reconsidered. In principle,
we expect a modest suppression of the $B_1'$ contribution.  We have
also neglected the possibility of $(B_1',B_1)$ mixing which could be sizable. 
It has been pointed out in Ref.~\cite{falk}
that in the $D$ meson system this indeed may occur.

There will be additional corrections coming from terms suppressed by
$m_B$ and by the chiral symmetry breaking scale $\L_\c$, which are
subleading terms in the heavy mass and chiral expansions respectively.
However we believe that one should not expect these or the  one loop
corrections to reduce the value of the couplings significantly, which
therefore will remain of order one or larger. In the case of
$\m_V$, the $B^*B\g$ coupling, the one loop corrections are always of
the oposite sign. However this is not always true for the $\m_A$'s,
where the corrections could even enhance the value of the coupling
considerably.

Perhaps the most important question in terms of the phenomenological
impact of these decays is the approximation made in Sec.~II, namely the
neglect of the form-factor suppression that affects $\m_V$ and
$\m_{A}^{(i)}$ and that would soften the spectrum. This peaks at a photon 
energy of $\approx 1.2$GeV when the couplings are approximated to be constant. 
When an energy suppresion is taking into account the average photon 
energy would drop typically to a few hundred MeV. 
 This suppression
would result in a smaller value of $R_{B}^{\m}$ by a factor that, as for the 
shape of the photon spectrum, 
strongly depends on the energy dependence chosen.  Typical energy
dependences would reduce $R_{B}^{\m}$ in (\ref{ratiomu}) by a factor of  $2$
to $4$. The first estimate is obtained with a monopole type suppression 
whereas the second case corresponds to an exponential suppression. In both 
cases the energy scale was chosen to be $1$GeV.    

On the other hand, we lack an understanding of the axial-vector meson
decay constants and there is a lot of room for theoretical
improvements. For the purpose of this letter, however, it was
sufficient to
show that a rough estimate of the plausible values of the relevant
couplings suggests that the branching ratio for the photonic decays
$B\to\m\n\g$ is comparable to $B\to\m\n$ and it could be even greater.

\section{Conclusions}
Radiative leptonic decays of heavy mesons are very interesting, not only 
as a possible background to purely leptonic decays, but also because they
yield new information about the strong and electromagnetic interactions of 
heavy hadrons. 
As expected, they  compensate for their
suppression by a factor of $\a$ by avoiding helicity suppression. In
the charmed mesons, $D$ and $D_s$ the effect is small for the muon
but sizeable for the electron.
  Allowing for a typical form-factor suppression
in $\m_V$ and $\m_{A}^{(i)}$ and assuming a large range for the
axial-vector decay constants we have
\bz
R_{D}^{\m}\approx (1-8)\times 10^{-2}\m_{V}^{2} \text{GeV}^2
\ez
which would translate into $Br(D\to\m\n\g)\approx 10^{-5}$.

The situation is very different for the $B$ meson decays. Since the
helicity suppression is so large, we benefit much more by
avoiding it; we get 
\bz
R_{B}^{\m}\approx (1-4)\m_{V}^{2}\left(  
1+\fr{1}{2}(\g_{1/2}+\g_{3/2})^2\right) \text{GeV}^2
\label{final}
\ez
We have seen in Sec.~II that the value of $\m_V$ in the $B$ mesons is
likely to remain a quantity  of order one even after considering
corrections that might reduce it. This indicates that a  measurement
of $f_B$ can only be achieved in the decays to muons if the radiative
decays are correctly substracted. 

This class of decays clearly deserves further study.
In particular it requires knowledge of the strong couplings $g$, $g'$,
$f'$ and $f''$ mentioned above. 
In any case it seems reasonable to expect values in the
range $R_{B}^{\m}=(1-20)$. This would translate into 
\bz
Br(B\to l\n\g)\approx (10^{-7}-10^{-6}) \nn
\ez
Therefore the observation of these decays
would be of great interest, adding to our understanding of $\m_V$,
$\m_{A}^{(i)}$ and the decay constants of orbitally excited heavy
mesons.

There are in principle several other decays of heavy mesons that
proceed via the SD mechanism.  For instance, the decay $B_s\to\m^+
\m^-\g$ will be enhanced over the helicity suppressed $B_s\to\m^+\m^-$
by an expression similar to (\ref{final}). In the Standard Model, the latter
has a branching ratio of $\approx 10^{-9}$ whereas the radiative decays
could be an order of magnitude greater. In  some theories beyond the SM 
these decays are expected to have larger branching ratios. For instance,
 it is argued in Ref.~\cite{etc} 
that Extended Technicolor scenarios enhance $Br(B_s\to\m^+\m^-)$ by up to two 
orders of magnitude, which would imply, together with (\ref{final}),  
$Br(B_s\to\m^+\mu^-\g)\approx 10^{-7}-10^{-6}$. This puts this radiative 
decay within experimental reach and, together with $B\to\m^+\m^-X$ 
\cite{gns}, can 
impose severe constraints on physics beyond the SM.

On the other hand, $B\to D^{(*)}\g l\n$ and
any other exclusive semileptonic decay with a photon  added to the
final state can be considered.
  Similar results may hold for non-leptonic decays.
Although these decays are indeed 
suppressed by $\a$ relative to 
the corresponding non-photonic decays, some of them are of interest 
in their own right and  might
became observable in the near future. 

\acknowledgments The authors thank J.F. Donoghue, A. Falk, Y. Kubota, C. Quigg 
and M. B. Wise for useful discussions and suggestions.
G.B. and T.G. acknowledge the support of the U.S. Department of 
Energy.

\noindent \figure{Fig.~1: Bremsstrahlung diagrams entering in  
$B\to\m\n\g$. The black squares denote the action of the weak 
current.\label{fig:1}}

\noindent \figure{Fig.~2: Structure Dependent diagrams entering in
$B\to\m\n\g$. The  black circles denote the action of the couplings $\m_V$
and $\m_{A}^{(i)}$. (a):~Vector meson contribution. (b):~Axial-Vector meson
contribution.  \label{fig:2}}

\noindent \figure{Fig.~3: One Loop corrections to $\m_V$ and  
$\m_{A}^{(i)}$. Dashed lines denote goldstone bosons while
 solid lines denote heavy mesons. \label{fig:3}}

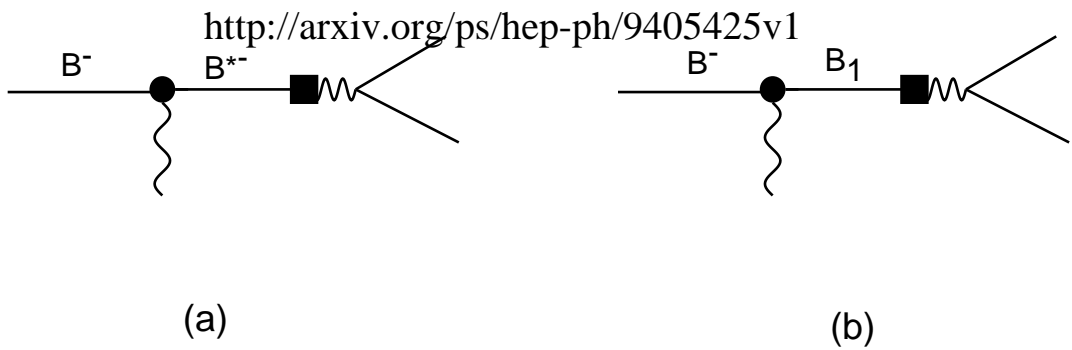
\end{document}